\documentclass[aps,prx,reprint,superscriptaddress,showpacs]{revtex4-1}
\usepackage{graphicx}
\usepackage{amsmath}
\usepackage{textcomp}
\usepackage{color}
\usepackage{xcolor}
\usepackage{upgreek}
\usepackage{comment}
\usepackage{verbatim}
\usepackage{amssymb}
\usepackage{tabu}
\usepackage{float}
\usepackage[english]{babel}
\usepackage{gensymb}
\usepackage{lineno}
\usepackage{mathrsfs}
\usepackage[pdfstartview={XYZ 0 0 1.0}]{hyperref} 

\begin{document}

\preprint{yes}

\title{Orbital-to-Spin Angular Momentum Conversion Employing Local Helicity}


\author{Sergey Nechayev}
\email[]{sergey.nechayev@mpl.mpg.de}
\affiliation{Max Planck Institute for the Science of Light, Staudtstr. 2, D-91058 Erlangen, Germany}
\affiliation{Institute of Optics, Information and Photonics, University Erlangen-Nuremberg, Staudtstr. 7/B2, D-91058 Erlangen, Germany}

\author{J{\"o}rg S.~Eismann}
\affiliation{Max Planck Institute for the Science of Light, Staudtstr. 2, D-91058 Erlangen, Germany}
\affiliation{Institute of Optics, Information and Photonics, University Erlangen-Nuremberg, Staudtstr. 7/B2, D-91058 Erlangen, Germany}

\author{Gerd Leuchs}
\affiliation{Max Planck Institute for the Science of Light, Staudtstr. 2, D-91058 Erlangen, Germany}
\affiliation{Institute of Optics, Information and Photonics, University Erlangen-Nuremberg, Staudtstr. 7/B2, D-91058 Erlangen, Germany}

\author{Peter Banzer}
\email[]{peter.banzer@mpl.mpg.de}
\homepage[]{http://www.mpl.mpg.de/}
\affiliation{Max Planck Institute for the Science of Light, Staudtstr. 2, D-91058 Erlangen, Germany}
\affiliation{Institute of Optics, Information and Photonics, University Erlangen-Nuremberg, Staudtstr. 7/B2, D-91058 Erlangen, Germany}


\date{\today}

\begin{abstract}
Spin-orbit interactions in optics traditionally describe an influence of the polarization degree of freedom of light on its spatial properties. The most prominent example is the generation of a spin-dependent optical vortex upon focusing or scattering of a circularly polarized plane-wave by a nanoparticle, converting spin to orbital angular momentum of light. Here, we present a mechanism of conversion of orbital-to-spin angular momentum of light upon scattering of a linearly polarized vortex beam by a spherical silicon nanoparticle. We show that focused linearly polarized Laguerre-Gaussian beams of first order ($\ell = \pm 1$) exhibit an $\ell$-dependent spatial distribution of helicity density in the focal volume. By using a dipolar scatterer the helicity density can be manipulated locally, while influencing globally the spin and orbital angular momentum of the beam. Specifically, the scattered light can be purely circularly polarized with the handedness depending on the orbital angular momentum of the incident beam. We corroborate our findings with theoretical calculations and an experimental demonstration. Our work sheds new light on the global and local properties of helicity conservation laws in electromagnetism.
\end{abstract}

\maketitle

\section{Introduction} 
Extensive research encompassing spin-orbit interaction (SOI)~\cite{Liberman1992,Bliokh2015} of light has been conducted to date owing to the fundamental importance and emerging nanophotonics applications~\cite{Bliokh2010, Aiello2015_11, Cardano2015} in a variety of fields, e.g. nanoparticle manipulation~\cite{Padgett2011}, directional coupling to spin-momentum-locked waveguide modes~\cite{Bliokh2011, Lee2012, Paco2013, Neugebauer2014, Petersen2014}, spin-controlled beam shaping~\cite{Yu2014}, spin-based photonics~\cite{Shitrit2013} and chiral quantum optics~\cite{Lodahl2017}, to name a few. 

SOI can be also observed in cylindrical symmetry, including, but not limited to, focusing of a beam by an aplanatic objective~\cite{Bomzon2006,Zhao2007,Nieminen2008}, scattering by a small particle~\cite{Schwartz2006,Haefner2009, Brasselet2009, Bliokh2011}, excitation and scattering of surface plasmon-polaritons~\cite{Gorodetski2008, Gorodetski2010, OConnor2014, Garoli2016} and transmission through a nanoaperture~\cite{Tischler2014, Zambrana2014, Garoli2016_07}. In cylindrical symmetry, the projection $J_z$ of the total angular momentum $\mathbf{J}$ of a beam on the axis of rotational symmetry $\mathbf{\hat{z}}$ is conserved~\cite{Bliokh2014}. Therefore, SOI in these systems typically manifests itself as a conversion of an incident spin angular momentum (SAM) to orbital angular momentum (OAM), that is, a generation of a spin-dependent optical vortex. 

Further insight into the physical origins of SOI can be obtained by considering an additional characteristic of an electromagnetic beam, i.e. the helicity $\sigma = \frac{\mathbf{J\cdot P}}{|\mathbf{P}|}$, which is defined as the projection of the total angular momentum $\mathbf{J}$ onto the direction of the linear momentum $\mathbf{P}$~\cite{Ivan2012, Bliokh2013, Ivan2013}. Importantly, $\sigma$ is the generator of the duality transformation~\cite{Ivan2013} and, hence, is preserved in systems and processes that posses duality symmetry, irrespective of the underlying geometry. Typical examples of dual-symmetric processes include scattering by dual scatterers~\cite{Zambrana2013, Nieto2015, Nieto2017}, propagation in piecewise-homogeneous impedance matched media~\cite{Li2009, Bliokh2010, Ivan2013} or focusing by an aplanatic objective designed to have equal Fresnel coefficients for $s$- and $p$-polarized incident beams \cite{Bliokh2010, Bliokh2011, Ivan2012}. Helicity is very intuitive in the far-field, where its density $K$ reduces to the proportion of circular polarization in each individual plane-wave component. Therefore, $\sigma$ in the far-field is the expression of the average SAM per plane-wave component~\cite{Ivan2012, Ivan2013, Bliokh2013, Nieto2015,helicity_note}. On the other hand, in real space, e.g. in the near field of a nanostructure or in the focal plane of a tightly focused beam, $K$ is more subtle because it originates from complex spatial distributions of three-dimensional fields $\mathbf{E,H}$~\cite{Tang2010, Bliokh2011_02, Bliokh2013, Bliokh2014_07}. Focused beams with zero far-field $K$ can show complex spatial distributions of $K$  in the focal plane. These peculiar far-field to near-field transformation properties of $K$ pave the way for performing \textit{local} operations on it to \textit{globally} affect $\sigma$ and SAM of the beam~\cite{manupulations_note,helicity_note}, similarly to operations on the $k$-space of a beam in the Fourier plane of a 4$f$ system to affect its spatial distribution. 

In this manuscript we employ local operations on $K$ to convert OAM of a linearly polarized beam to SAM. Firstly, we show that a focused linearly polarized Laguerre-Gaussian beam of first order $\left(\ell = \pm 1 \right)$ (from this point onwards referred to as $\text{LG}_{\pm 1}$)~\cite{Allen1992} exhibits $\ell$-dependent values of $K$ in the focal plane. Secondly, we utilize a dipolar Mie-scatterer~\cite{bohren1983} positioned on the optical axis in the focal volume to manipulate $K$ of such a beam locally. As a consequence, we obtain two distinct regimes of OAM to SAM conversion. For the first regime we consider a scatterer, which is dual-symmetric at a particular wavelength $\lambda_d$~\cite{Zambrana2013}. 
Because a dipolar scatterer responds only to the local helicity density $K$ of the beam, and not to its integrated zero value, we can show that the dipole moment excited in the nanoparticle at $\lambda_d$ emits purely circularly polarized light with a handedness defined by $K$ and, eventually, by the OAM of the incident beam. Even though this results in SAM in the far-field, since the scatterer is dual-symmetric at $\lambda_d$, we observe no total generation of helicity~\cite{Ivan2013, Nieto2015, helicity_note}. Nevertheless, this is different for the second regime at a wavelength $\lambda\neq\lambda_d$, with the scatterer breaking the duality symmetry. In this case the nanostructure locally extincts helicity from an initially linearly polarized beam~\cite{Nieto2017, Nieto2017_05}, resulting in a total generation of far-field helicity~\cite{helicity_note}. We treat the aforementioned cases theoretically, and demonstrate experimentally the conversion of OAM to SAM by a dual-symmetric dipole scatterer~\cite{Zambrana2013_07}.

\section{Theory} 
\subsection{Reflected and transmitted far-fields} 

\begin{figure}
  \includegraphics[width=0.48\textwidth]{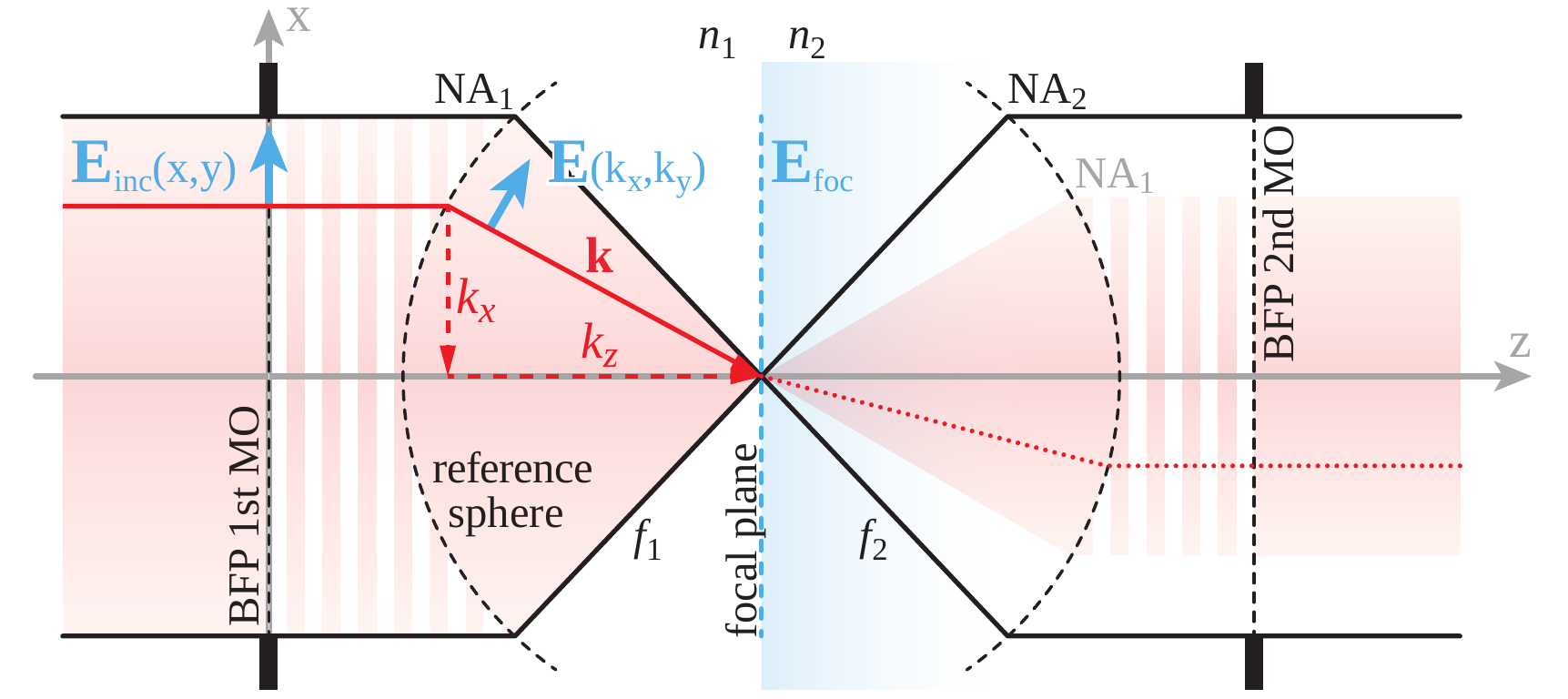}
  \caption{Simplified sketch of the investigated system for parameters $n_1<n_2$ and $\text{NA}_1<\text{NA}_2$.	An aplanatic high numerical aperture (NA) system is used to tightly focus an incoming beam impinging from left to right. The incoming electromagnetic field $\mathbf{E}_{\text{inc}}$ at the back focal plane (BFP) of the first microscope objective (MO) is projected onto a reference sphere with radius $f_1$. A second confocally aligned aplanatic system is used to collect the transmitted light in the second half-space.}
  \label{fig:focusing}
\end{figure}

We start by briefly introducing the investigated scheme shown in Fig.~\ref{fig:focusing}. We use two confocally aligned microscope objectives (MO) with focal length $f_i$, described as aplanatic systems, where the focal plane of our system separates the left half-space ($z<0$, $i=1$) and the right half-space ($z>0$, $i=2$). Both half spaces are non-absorbing non-magnetic dielectrics characterized by their refractive index $n_i=\sqrt{\varepsilon_i \mu_i}$ ($\varepsilon_i$ and $\mu_i=1$ are the relative permittivity and permeability, respectively)  and the numerical aperture NA$_i$ of their aplanatic system. Each of the two MO's is index-matched to the refractive index of its corresponding half-space. In this system, the first MO focuses the incoming beam and collects the reflected light from the optical boundary at the focal plane, whereas the transmitted light is collected by the second MO. 

We consider a paraxial $x$-polarized $\text{LG}_{\pm 1}$ beam $\mathbf{E}_{\text{in}}=E_{\text{in}} \left(x,y\right)\hat{\mathbf{x}}=E_0\frac{\rho}{w_0}\exp\left(- \frac{\rho^2}{w_0^2}+\imath \ell \varphi\right) \hat{\mathbf{x}}$, illuminating the back focal plane (BFP) of the first MO, where $\rho=\sqrt{x^2+y^2}$ and $\varphi=\arctan\left(y/x\right)$ are the radial and axial cylindrical coordinates. Furthermore, $w_0$ is the beam waist and $\ell = \pm 1$ is the topological charge of the incoming beam. Following the approach described in ref.~\cite{novotny2006}, the field distribution at the entrance aperture of the MO can be linked to $k$-space via the transverse Cartesian coordinates: $x=-f_1\frac{k_x}{k_1}, y=-f_1\frac{k_y}{k_1}$, where $k_i=k_0 n_i$ is the wavenumber of the corresponding half-space and $k_0$ is the free-space wavenumber.
The highest transverse $k$-vector, which can be focused or collected by our aplanatic systems, is defined by the corresponding numerical aperture NA$_i \geq k_\bot/k_0$ and given by $k_\bot=\sqrt{k_x^2+k_y^2}$. The transmitted fields in the BFP of the second MO $\left(\mathbf{E}^{\infty}_t\right)$ and the field distributions of the reflected fields in the BFP of the first MO $\left(\mathbf{E}^{\infty}_r\right)$  can be written as:

\begin{align} \begin{split}
\left[\begin{matrix} E^{\infty}_{t,p}\\ E^{\infty}_{t,s} \end{matrix}\right]
(k_{x},k_{y})&=\frac{O_1}{O_2} 
\left[\begin{matrix} ~~k_x t_p/k_\bot\\ -k_y t_s/k_\bot\\ \end{matrix}\right]
E_{\text{in}}(k_{x},k_{y}), \\
\left[\begin{matrix} E^{\infty}_{r,p}\\ E^{\infty}_{r,s} \end{matrix}\right]
(k_{x},k_{y})&=
\left[\begin{matrix} k_x r_p/k_\bot\\  k_y r_s/k_\bot\\ \end{matrix}\right]
E_{\text{in}}(k_{x},k_{y}), 
\end{split} \label{eq:ff} \end{align}
respectively. Here, $\mathrm{E}^{\infty}_{p}$ and $\mathrm{E}^{\infty}_{s}$ are the radial and azimuthal field components in a cylindrical reference frame and $t_p,~t_s,~r_p,~r_s$ are the corresponding Fresnel transmission and reflection coefficients \cite{novotny2006}, respectively. Last, the factors
\begin{equation}
O_i=\frac{\imath f_i \exp\left( - \imath k_i f_i \right) }{2 \pi \sqrt{k_{z_i}k_i }}, \nonumber
\label{eq:stationary_phase}
\end{equation}
link the far-field on the reference sphere with the $k$-spectrum of the electric field via the method of stationary phase, which is described in a detailed manner in chap.~3.3 in~\cite{mandel_wolf}. Additionally, $k_{z_i}=\sqrt{k_i^2-k_\perp^2}$ is the longitudinal component of $k_i$ with $\Im(k_{z_i})>0$.

\subsection{Focal fields and helicity decomposition} 
\label{chap:focal_fields_and_helicity_decomposition}
\begin{figure*}
  \includegraphics[width=\textwidth]{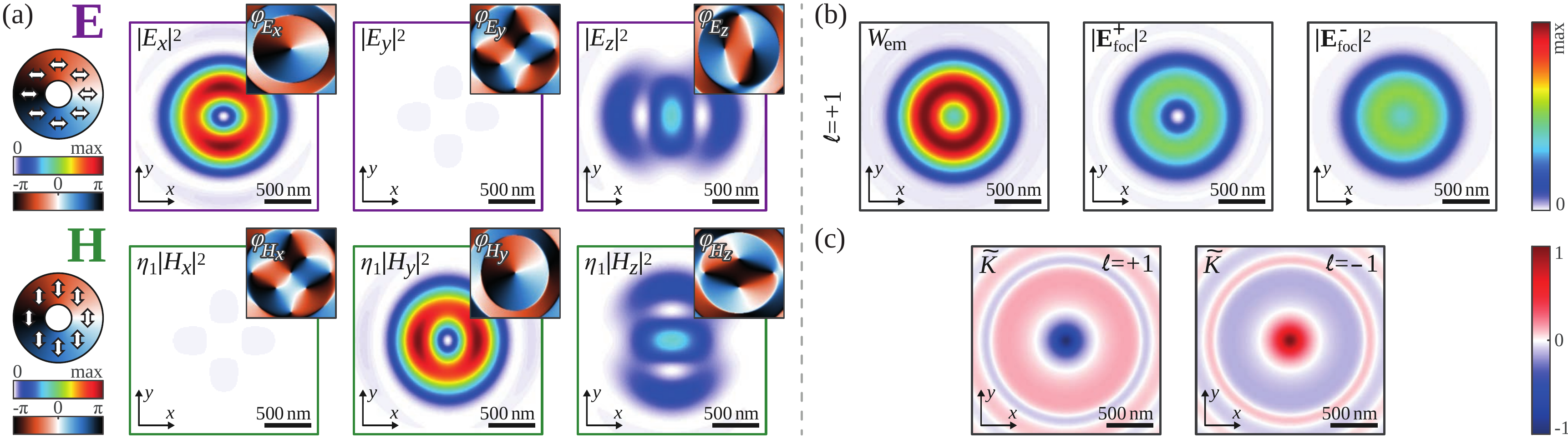}
  \caption{ Calculated properties of an $x$-polarized $\mathrm{LG}_{\pm1}$ beam at a wavelength of $\lambda=715$\,nm, tightly focused in free-space ($n_1=n_2=1$) with a numerical aperture of 0.9.
	(a) Electric and magnetic focal-field distributions for $\ell=+1$ with their corresponding phases shown as insets in the top right corners. The polarization distribution of the electric and magnetic field in the paraxial regime is shown on the left. 
	(b) Energy densities for the same beam as in (a). $\mathcal{W}_{\text{em}}$ shows the total electromagnetic energy density, whereas $|\mathbf{E}_{\text{foc}}^+|^2$ and $|\mathbf{E}_{\text{foc}}^-|^2$ present the electric energy density for only LCP and RCP components of the angular spectrum, respectively. On the optical axis only RCP components contribute to $\mathcal{W}_{\text{em}}$.
	(c) Spatial distributions of the normalized helicity density $\widetilde{K}= k_0K/\mathcal{W}_{\text{em}}$ in the focal plane for an incident beam with $\ell=+1$ and $\ell=-1$. }
  \label{fig:focal_fields}
\end{figure*}

Utilizing the plane wave decomposition explained in the previous chapter, it is also possible to calculate the focal field distributions of an arbitrary input beam~\cite{novotny2006}. For the case of focusing in freespace ($n_1=n_2=1$, no reflection), we show the calculated focal fields of a $x$-polarized $\mathrm{LG}_{+1}$ beam in Fig.~\ref{fig:focal_fields} (a). Adapted to our experimental situation described later, we use a focusing objective with a numerical aperture of NA$_1$=0.9 and an aperture filling factor of $\frac{w_0}{f_1\text{NA}_1}=0.71$ at a wavelength of $\lambda= 715\,$nm for calculations.

As we see from the focal field distributions, only the longitudinal field components are present on the optical axis in the focal plane, which satisfy $H_z=\imath \ell / \eta_1^{-1} E_z$, where $\eta_1=\sqrt{\mu_1\mu_0/\varepsilon_1 \varepsilon_0}$ is the impedance of the medium. The phase difference $\Delta \phi$ between $H_z$ and $E_z$ gives rise to a helicity density $K=- \left( \varepsilon_1 \varepsilon_0 \eta_1 / 2k_0 \right) \Im \left( \mathbf{E}^\ast\cdot\mathbf{H}\right)$~\cite{Tang2010, Bliokh2013, Bliokh2011_02, Bliokh2014_07}. Even though $K$ in the focal volume depends on both $\mathbf{E}$ and $\mathbf{H}$, it can be derived from the electric field components only taking advantage of the so-called helicity basis representation~\cite{Berry2009, Bliokh2011_02,Bekshaev2011_03, Ivan2012, Aiello2015, Nieto2017}. This decomposition allows for separate discussion on the contributions of LCP and RCP polarized components of electric and magnetic fields as follows. To proceed we first decompose the incident paraxial beam into its circularly polarized components:
\begin{equation}
\mathbf{E}_{\text{in}}=
\frac{1}{\sqrt{2}}E_{\text{in}} \left[ \frac{\left(\hat{\mathbf{x}} + \imath \hat{\mathbf{y}} \right)}{\sqrt{2}} +\frac{\left(\hat{\mathbf{x}} - \imath \hat{\mathbf{y}} \right)}{\sqrt{2}} \right]
\equiv \mathbf{E}_{\text{in}}^+ +\mathbf{E}_{\text{in}}^-,
\end{equation} 
where $\mathbf{E}_{\text{in}}^+,\, \mathbf{E}_{\text{in}}^-$ are the LCP and RCP polarized components, respectively. Next, the focal fields for each of the components $\mathbf{E}_{\text{foc}}^+,\, \mathbf{E}_{\text{foc}}^-$ are calculated independently. It can be shown~\cite{Berry2009, Bliokh2011_02,Bekshaev2011_03, Ivan2012, Aiello2015, Nieto2017} that the total electric and magnetic focal fields are given by $\mathbf{E}_{\text{foc}}=\mathbf{E}_{\text{foc}}^+ + \mathbf{E}_{\text{foc}}^-$ and $\mathbf{H}_{\text{foc}}=-\imath \eta_1^{-1} \left[\mathbf{E}_{\text{foc}}^+-\mathbf{E}_{\text{foc}}^-\right]$. As a consequence, the total electric \textit{and} magnetic energy density can be expressed as a sum of the contributions originating form LCP and RCP electric field components $\mathcal{W}_{\text{em}} =\left( \varepsilon_1\varepsilon_0 /2 \right)\left[| \mathbf{E}^+ |^2 +|\mathbf{E}^- |^2 \right]$. Furthermore, because $\mathbf{E}^+,\, \mathbf{E}^-$ only include contributions of LCP and RCP plane-waves, respectively, the helicity density $K$ in the focal volume and in the far-field is proportional to a difference between these contributions $K= \left( \varepsilon_1 \varepsilon_0 / 2k_0 \right) \left[ | \mathbf{E}^+ |^2 -|\mathbf{E}^- |^2 \right]$. 

In Fig.~\ref{fig:focal_fields} (b) we plot the total energy density $\mathcal{W}_{\text{em}}$ in the focal plane as well as the components $\mathbf{E}_{\text{foc}}^+$ and $\mathbf{E}_{\text{foc}}^-$.
We can see that $\mathbf{E}_{\text{foc}}^+$ is zero on the optical axis, whereas $\mathbf{E}_{\text{foc}}^-$ shows a significant energy density at this point~\cite{Zambrana2014}. The reason for the qualitatively different spatial distributions of $\mathbf{E}_{\text{foc}}^+$ and $\mathbf{E}_{\text{foc}}^-$ is the spin-to-orbit angular momentum conversion upon focusing and the different total angular momenta in $\mathbf{E}_{\text{in}}^+$ and $\mathbf{E}_{\text{in}}^-$~\cite{Allen1992, Bliokh2010, Bliokh2011, Bliokh2014}. Additionally, in Fig.~\ref{fig:focal_fields} (c) we show that the values of the normalized helicity density $\widetilde{K}= k_0K/\mathcal{W}_{\text{em}} \in [-1,1]$, for an incident beam with $\ell = \pm 1$ exhibit maximum absolute values on the optical axis ($\widetilde{K} = -\ell$)~\cite{Bliokh2011_02, Bliokh2014_07}. These values are equal to those obtained for an RCP or LCP plane-wave, respectively. This is consistent with Fig.~\ref{fig:focal_fields} (b), showing that in the focal plane on the optical axis only RCP ($\ell = +1$) plane-waves contribute to the focal fields.

\subsection{Far-field scattered light and orbit-to-spin coupling} 
\label{chap:scatterer}

As a next step, we now assume that the 3D focal fields excite a dipolar high refractive-index dielectric scatterer, positioned on the optical axis at $\mathbf{r}_0=(0,0,-d)$ with $d>0$. The scatterer is characterized by its first order electric and magnetic Mie coefficients $a_1(\lambda)$ and $b_1(\lambda)$, which are complex functions of the wavelength $\lambda$~\cite{bohren1983, Hightower1988}. The induced electric and magnetic dipole moments can then be calculated by $\mathbf{p}=6 \pi \imath \varepsilon_0 n_1^2 / k_1^3 a_1 \mathbf{E}_{\text{foc}}(\mathbf{r}_0)$ and  $\mathbf{m}= 6 \pi \imath / k_1^3 b_1 \mathbf{H}_{\text{foc}}(\mathbf{r}_0)$, where $\varepsilon_0$ is the vacuum permittivity. Consequently, when the scatterer is placed on-axis, only longitudinal electric and magnetic dipole moments $p_z$ and $m_z$ can be excited, owing to the focal field distributions of the chosen beam [see Fig.~\ref{fig:focal_fields} (a)]. Therefore, the far-field scattered light in the BFP of the MOs acquires a simple form in cylindrical coordinates:
\begin{align} \begin{split}
\left[ \begin{matrix} E^{d}_{t,p}\\ E^{d}_{t,s} \end{matrix}\right]
(k_{x},k_{y})&=\frac{C}{O_2} D
\left[ \begin{matrix} \frac{ -k_\bot p_z}{k_1} t_p   \\ \frac{k_\bot m_z}{c_1 k_1} t_s \\ \end{matrix} \right], \\
\left[ \begin{matrix} E^{d}_{r,p}\\ E^{d}_{r,s} \end{matrix}\right]
(k_{x},k_{y})&=\frac{C}{O_1}
\left[ \begin{matrix} \frac{ -k_\bot p_z}{k_1} \left( \frac{1}{D} +Dr_p \right) \\ \frac{k_\bot m_z}{c_1 k_1} \left(\frac{1}{D} + Dr_s\right) \\ \end{matrix}\right]. 
\end{split} \label{eq:ff_dipoles} \end{align}
Here, $\mathbf{E}^{d}_t$ contains the forward scattered and transmitted light, while $\mathbf{E}^{d}_r$ describes the backward scattered as well as the forward scattered but reflected parts. In addition, $C=\frac{\imath k_0^2}{8 \pi^2 \varepsilon_0 k_{z_1}}$, $D=\exp \left( \imath k_{z_1}d \right)$ and $c_i$ is the speed of light in medium $i$. The total electric field in the BFP of the second MO ($\mathbf{E}_t$) and the first MO ($\mathbf{E}_r$) can be obtained by summing Eq.~\eqref{eq:ff} and \eqref{eq:ff_dipoles}:
\begin{align} \begin{split}
\mathbf{E}_t\left(k_{x},k_{y}\right)=&
\left[\begin{matrix} E^{\infty}_{t,p}\\ E^{\infty}_{t,s} \end{matrix}\right] +
\left[ \begin{matrix} E^{d}_{t,p}\\ E^{d}_{t,s} \end{matrix}\right], \\
\mathbf{E}_r\left(k_{x},k_{y}\right)=&
\left[\begin{matrix} E^{\infty}_{r,p}\\ E^{\infty}_{r,s} \end{matrix}\right] +
\left[ \begin{matrix} E^{d}_{r,p}\\ E^{d}_{r,s} \end{matrix}\right]. 
\end{split} \label{eq:ff_total} \end{align}

\begin{figure}
  \includegraphics[width=0.48\textwidth]{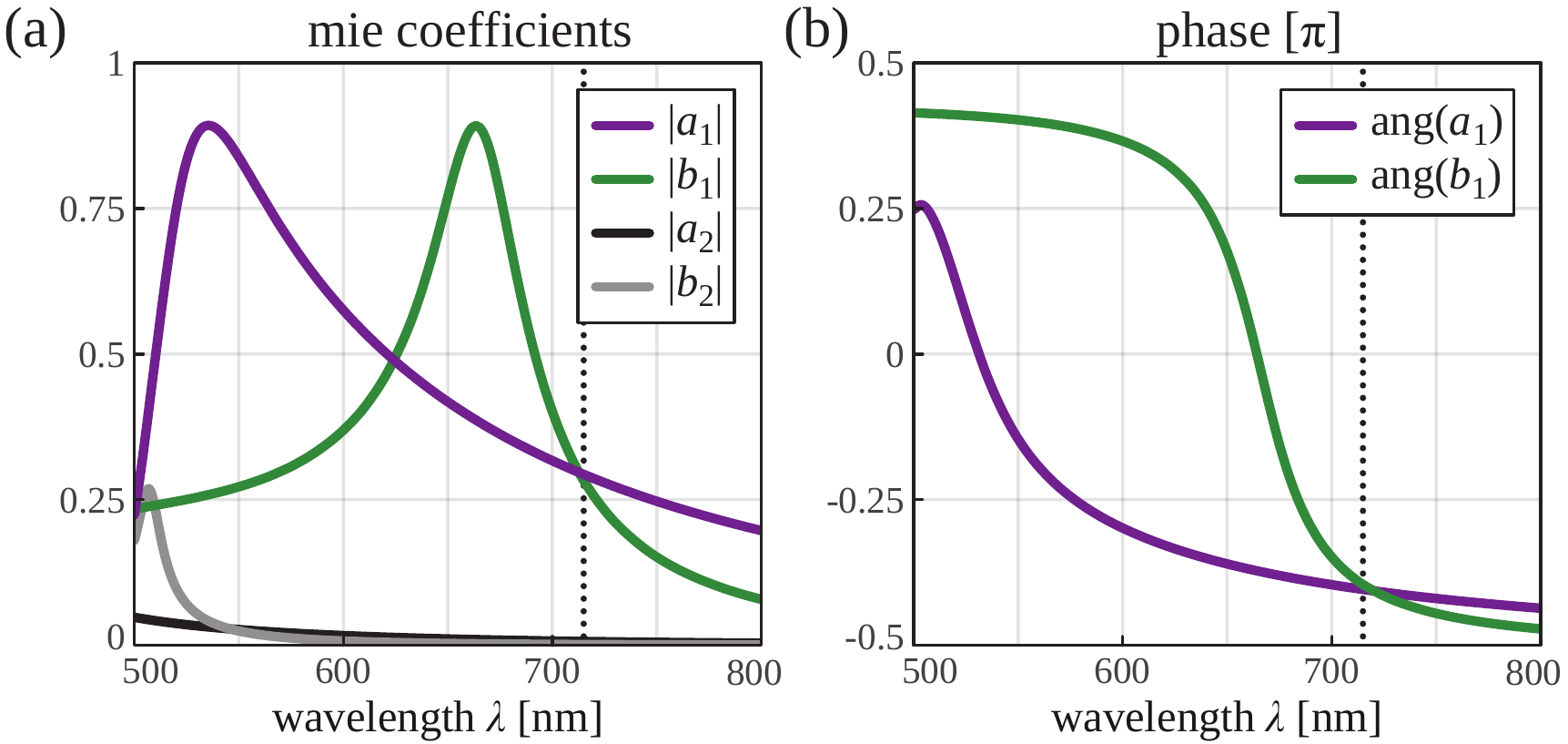}
  \caption{(a) Absolute values of the first and second order Mie coefficients of a core-shell nanoparticle with Si core of radius $r_{\text{Si}} = 83$\,nm and a SiO$_2$ shell of thickness $\delta = 4$\,nm. (b) Corresponding phases of the first order Mie coefficients. The dotted black lines show the wavelength $\lambda_d= 715$\,nm where the particle is approximately dual-symmetric, i.e. $a_1\approx b_1$.}
  \label{fig:mie}
\end{figure}

Inspired by the scattering particle utilized later in the experiment, from this point onwards the scatterer will be a spherical concentric core-shell nanosphere at the position $\mathbf{r}_0=(0,0,-87\text{\,nm})$. The core of the nanoparticle features a radius of $r_{\text{Si}} = 83$\,nm and consists of crystalline silicon~\cite{Palik1985}, whereas the shell material is SiO$_2$~\cite{Palik1985} with an estimated thickness of $\delta= 4$\,nm~\cite{shell,Decker2016}. In Fig.~\ref{fig:mie} (a) and (b) we plot the first and second order Mie coefficients~\cite{bohren1983, Hightower1988} and their corresponding phases. There we can see that for a wavelength $\lambda \geq 600$\,nm, the first order Mie coefficients $a_1$, $b_1$ are sufficient to characterize the scatterer. Moreover, at the wavelength $\lambda_{\text{d}} \approx 715$\,nm the first Kerker condition~\cite{Kerker1983, Geffrin2012, Zambrana2013, Zambrana2013_07} is approximately satisfied, i.e.~$a_1=b_1$, as marked by a dotted black line in Fig.~\ref{fig:mie} (a) and (b). Additionally, for a homogeneous medium ($n_1=n_2$), the condition for electric and magnetic fields exciting the particle at the focal point $H_z=\imath \ell \eta_1^{-1}E_z$, is also fulfilled on the optical axis outside of the focal plane (as a result also $\widetilde{K} = -\ell$ is fulfilled there).
Thus, the excited dipole moments $p_z$ and $m_z$ fulfill $m_z =\imath \ell c_1 p_z$. This combination of parallel electric and magnetic dipoles phase shifted by $\pm \pi/2$ has been termed $\sigma$-dipole~\cite{Zambrana2016, Eismann2018} since in free-space it emits light with a well-defined helicity $\sigma$ of $\pm1$ in all directions. In order to prove the pure circular polarization in the far-field, we insert the relation between the excited electric and magnetic $z$-dipoles into Eq.~\eqref{eq:ff_dipoles} and obtain $E^{d}_p = \imath \ell E^{d}_s$, for all $(k_x,k_y)$ in forward as well as in backward direction. This relation between the $p$- and $s$-polarized electric field components confirms that the scattered far-field is circularly polarized, with a handedness depending on the sign of the OAM of the incoming LG beam. In particular, the emitted light is purely RCP polarized for the case of $\ell=+1 $ and LCP polarized for $\ell=-1$~[cf.~Fig.~\ref{fig:focal_fields} (b) and (c)]. It is also worth mentioning that in some directions, e.g. backwards, the scattered light does not interfere with the incident beam for a particle in free-space, which keeps the far-field purely circularly polarized for those angular regions~\cite{helicity_note}. 

So far, we have presented theoretically a way to employ the helicity density $K$ to convert OAM of the incident linearly polarized light to SAM of the scattered light (orbit-to-spin coupling). 

\subsection{Helicity conservation} 

The circularly polarized light emission from the excited dipole moment discussed in the previous section provides a very deep insight into global and local properties of recently derived theorems of conservation of helicity~\cite{Ivan2012, Ivan2013, Nieto2015} and the role of duality symmetry in optics. At the chosen wavelength $\lambda_d=715\,\mathrm{nm}$ the scatterer is approximately dual-symmetric, hence featuring interesting properties.

\begin{figure}
  \includegraphics[width=0.48\textwidth]{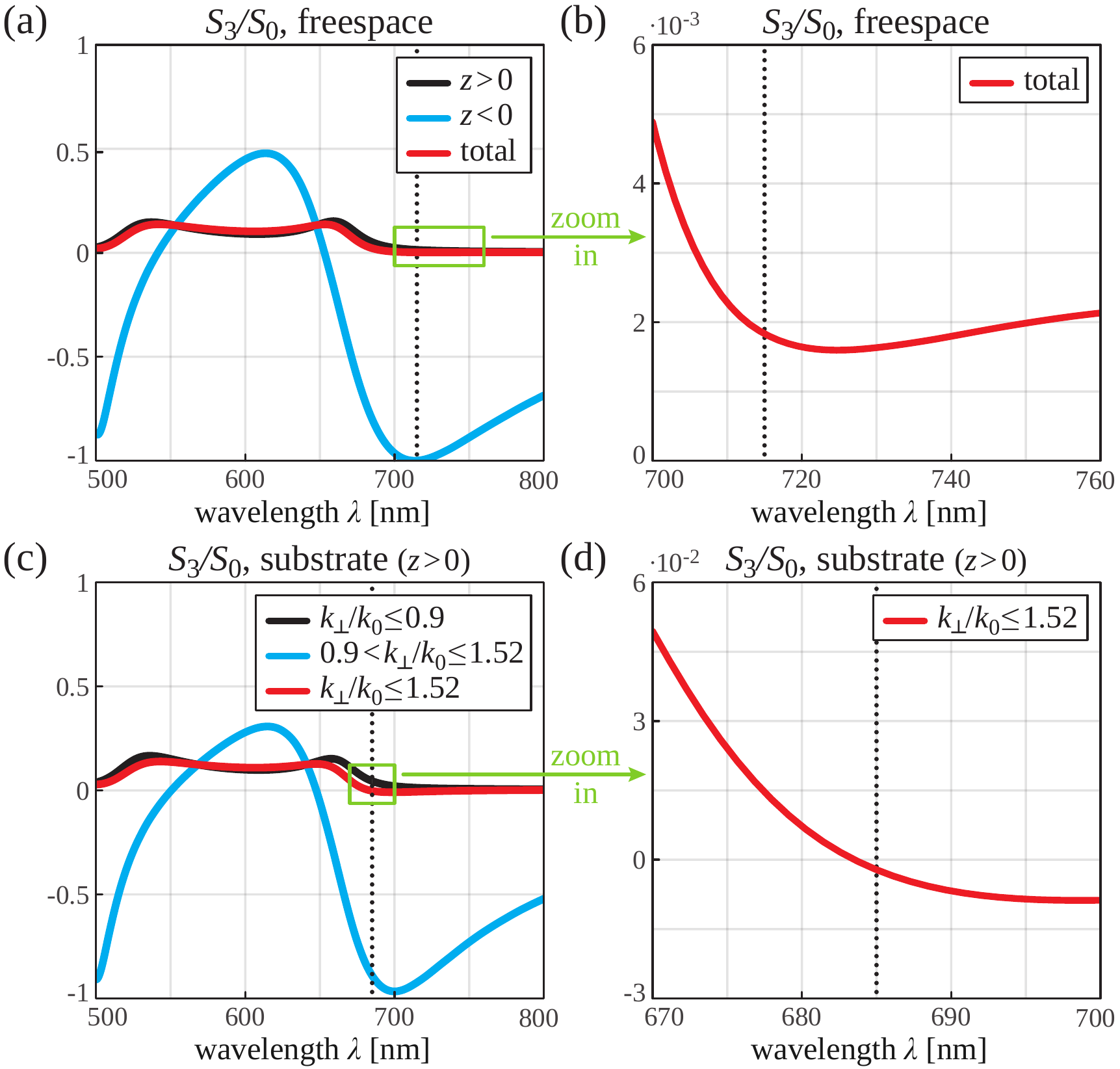}
  \caption{ 
Scattering of a tightly focused $\mathrm{LG}_{+1}$ beam by the nanoparticle shown in Fig.~\ref{fig:mie}. 
We show the average helicity ($S_3/S_0$) of the emitted light by plotting the integrated Stokes $S_3$ parameter normalized by the integrated $S_0$ parameter. The wavelength $\lambda_d$ where the particle is approximately dual-symmetric is indicated by dotted black lines. For free-space ($n_1=n_2=1$), (a) and (b) present the left half-space (blue), the right half-space (black) and the total value (red), where (b) shows an enlarged view onto the total value around $\lambda_d$. (c) and (d) show the case of the same scatterer positioned on a dielectric substrate ($n_1=1,\,n_2=1.52$), where we integrate over the light emitted in forward direction over different angular regions. The blue, black and red curve show $S_3/S_0$ for $0.9<\text{NA}_2 \leq 1.52$, $\text{NA}_2 \leq0.9$ and $\text{NA}_2 \leq 1.52$, respectively. (d) shows the area around $\lambda_d$ in more detail.
} \label{fig:helicity}
\end{figure}

At first, a dual-symmetric scatterer has to preserve the local helicity. Therefore, the scattered light helicity is defined by the local helicity-density of the excitation field at the position $\mathbf{r}_0$ of the particle. Consequently, to show the response to the local helicity density upon scattering by a dual-symmetric dipolar particle, we integrate the resulting far-field Stokes parameter $S_3$ in backward direction in regions of no interference with the excitation field, and normalize it by the integrated far-field total Stokes parameter $S_0$ in the same angular region. In Fig.~\ref{fig:helicity} (a), we plot the resulting spectrum of $S_3/S_0$ calculated with Eqns.~\eqref{eq:ff_total} for our scatterer (Fig.~\ref{fig:mie}) and excitation beam in free-space [Fig.~\ref{fig:focal_fields}(a)]. The results are shown for the backward scattered light ($z<0$, blue line), the light propagating in forward direction ($z>0$, black line) and in full solid angle (red line). A close look at the blue curve confirms the response to the local helicity density, because at $\lambda_d$ the light scattered in backwards direction is purely RCP polarized. This is consistent with our calculations presented in Fig.~\ref{fig:focal_fields} (c), where we saw $\widetilde{K}=-1$ for $\ell=+1$ on the optical axis.

Owing to helicity conservation theorems for dual non-absorbing scatterers~\cite{Ivan2012, Ivan2013, Nieto2015}, also the global helicity of the interference between incident and scattered light must be equal to that of the incident field featuring zero helicity. At $\lambda_d$, $S_3/S_0$ integrated over full solid angle must be approximately zero (see Fig.~\ref{fig:helicity} (a) and (b)). In Fig.~\ref{fig:helicity} (b), we see a value close to zero, red-shifted with respect to $\lambda_d$, since for the scatterer, even if it was lossless, $a_1\approx b_1$, but $a_1\neq b_1$.

In Addition to the conservation of the total helicity at $\lambda_d$ and the generation of SAM along the propagation direction $\mathbf{\hat{z}}$, the red curve in Fig.~\ref{fig:helicity} (a) also reveals that when using an excitation wavelength that causes our scatterer to break the dual symmetry, it is also possible to globally convert OAM into helicity. In this manner a dipolar spherical (and achiral) scatterer performs an operation on $K$ in a cylindrically symmetric system in a way that locally extincts helicity~\cite{Nieto2017, Nieto2017_05} in the focal plane of the initially linearly polarized beam, resulting in a total generation of helicity in the far-field. Hence, this regime corresponds to the average conversion of OAM to helicity for non dual-symmetric conditions~\cite{helicity_note}. 

To experimentally confirm orbit-to-spin conversion, the backward scattered light has to be collected and analyzed for a homogeneously embedded particle. Alternatively, we can place the scatterer on a higher-index dielectric substrate, which facilitates the demonstration of orbit-to-spin conversion in two ways. Firstly, the backward scattering is strongly suppressed~\cite{novotny2006} and most of the light emitted by the nanoparticle is coupled to forward direction. Secondly, in the supercritical angular region (above the critical angle, $k_{\bot}>k_1$), only scattered light is observable creating an angular region without interference with the incident beam. We therefore expect the light emitted to the supercritical region to be almost purely RCP polarized at wavelengths close to $\lambda_d$. We calculate and integrate $S_3$ and $S_0$ by using Eq.~\eqref{eq:ff_total} for the particle presented in Fig.~\ref{fig:mie} positioned in air on a glass substrate ($n_1=1, n_2=1.52$). In our calculations the scatterer is excited by the focused incident and reflected field, while the excitation by the reflected scattered light is neglected. In Fig.~\ref{fig:helicity} (c), we show $S_3/S_0$ for different angular regions in forward direction --- $k_\bot/k_0 \leq 0.9 = \text{NA}_1$ (black), $0.9 < k_\bot/k_0 \leq 1.52$ (blue) and $k_\bot/k_0 \leq 1.52 = \text{NA}_2$ (red). Since in the region above the NA of the focusing objective only scattered light is present, the blue curve in Fig.~\ref{fig:helicity} (c) resembles the blue one in (a). However, the minimum is blue shifted by approximately 15\,nm, since the substrate influences the effective polarizability of the nanoparticle. Moreover, the minimum does not reach the value of minus one, owing to the complex nature of the Fresnel coefficients in the supercritical angular region. In Fig.~\ref{fig:helicity} (d), which shows a magnified area from (c), we observe that the average helicity in forward direction crosses zero at a wavelength of $\lambda_{d,s}=685$\,nm. This is the wavelength that we will use for an experimental demonstration later on. Although $\lambda_{d,s}$ does not correspond to the minimum of the blue curve in Fig.~\ref{fig:helicity} (c), the scattered light will still be strongly circularly polarized. In addition, since the dipole moments excited in the nanoparticle are oscillating along the substrate normal, most of the scattered light will be emitted to a narrow angular region around the critical angle~\cite{novotny2006}, facilitating the experimental observation.

\section{Experimental realization} 
\begin{figure}
  \includegraphics[width=0.48\textwidth]{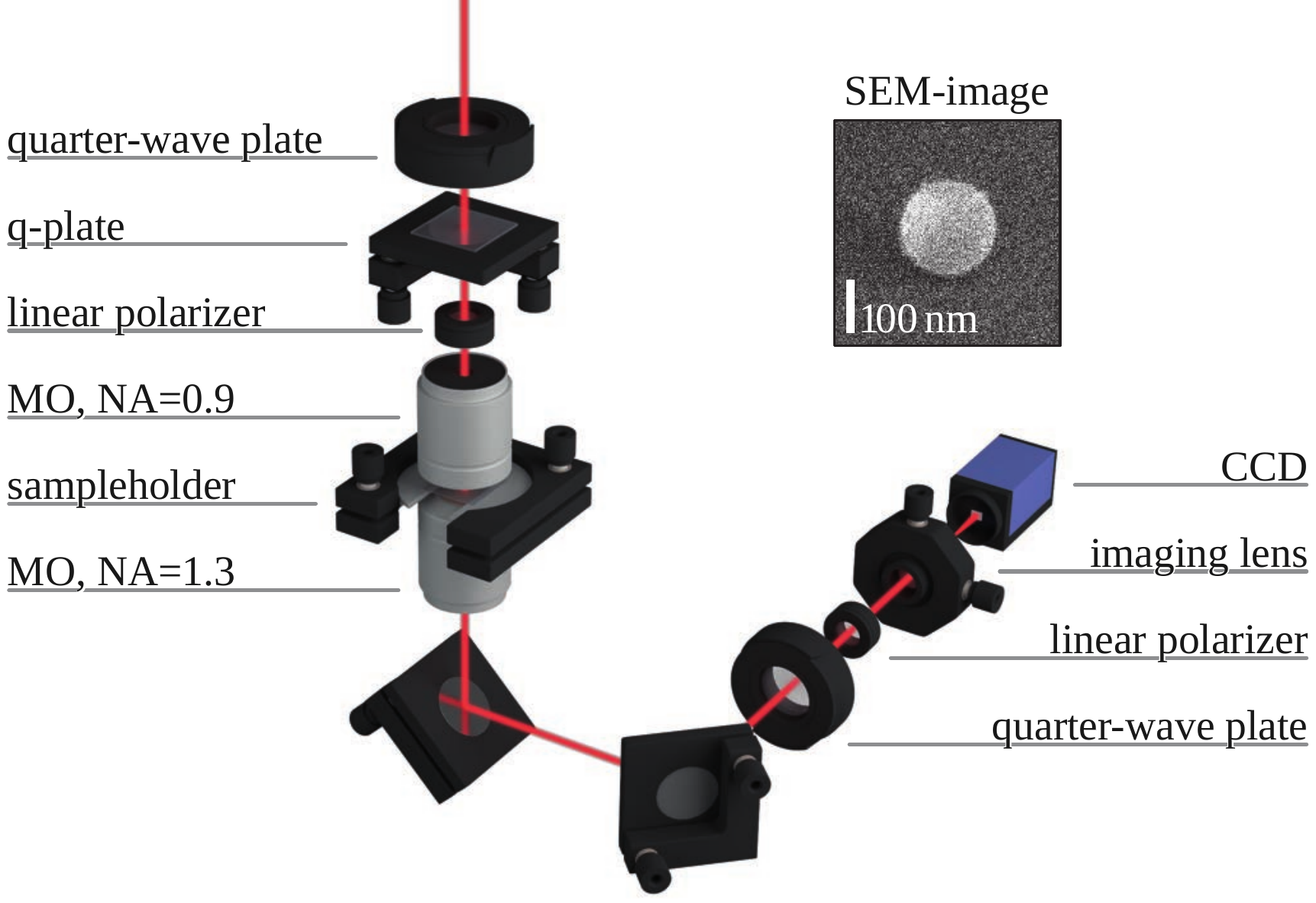}
  \caption{Sketch of the experimental setup. 
	A quarter wave plate, a q-plate of charge $-1/2$ and a linear polarizer transform the incoming linearly polarized Gaussian beam into an $\text{LG}_{\pm 1}$ mode. The paraxial beam is tightly focused onto a silicon nanoparticle (radius=87\,nm, SEM-image shown as inset) by a microscope objective (MO). The light propagating in forward direction is collected by an immersion-type MO. A rotatable quarter-wave plate and a linear polarizer are utilized for polarization analysis before a lens images the back focal plane of the second MO onto a CCD-camera.}
  \label{fig:setup}
\end{figure}

The main part of the experimental setup, which is similar to that presented in previous works \cite{Banzer2010, Eismann2018}, is shown as a simplified sketch in Fig.~\ref{fig:setup}. An incoming linearly polarized Gaussian beam with a wavelength of 685\,nm is converted into an $\text{LG}_{\pm 1}$ beam by the use of a quarter-wave plate, a q-plate~\cite{Marrucci2006} of charge $-1/2$ and a linear polarizer. The sign of the charge $\ell$ of the generated LG beam can be set by aligning the axis of the quarter-wave plate with an angle of $\pm45^\circ$ relative to the incoming linear polarization. Afterwards, the beam is tightly focused by the first MO with NA$_1$=0.9 onto a silicon nanoparticle sitting on a glass substrate. An SEM-image of the particle with a radius of 87\,nm is shown as an inset in Fig.~\ref{fig:setup}. Precise positioning of the particle with respect to the beam is enabled by a 3D-piezo stage, attached to the substrate. Utilizing an index matched oil immersion MO (NA$_2=1.3$) in a confocal alignment with the first MO, the beam transmitted through the interface as well as the light scattered by the particle is collected and collimated. In order to measure in the far-field of our system, we image the BFP of the second MO onto a CCD camera. Prior to the imaging lens, a rotatable quarter-wave plate together with a linear polarizer are placed to project the light onto different polarization states, enabling us to reconstruct the far-field Stokes parameters \cite{Schaefer2007}.

\section{Results and discussion} 

\begin{figure}
  \includegraphics[width=0.48\textwidth]{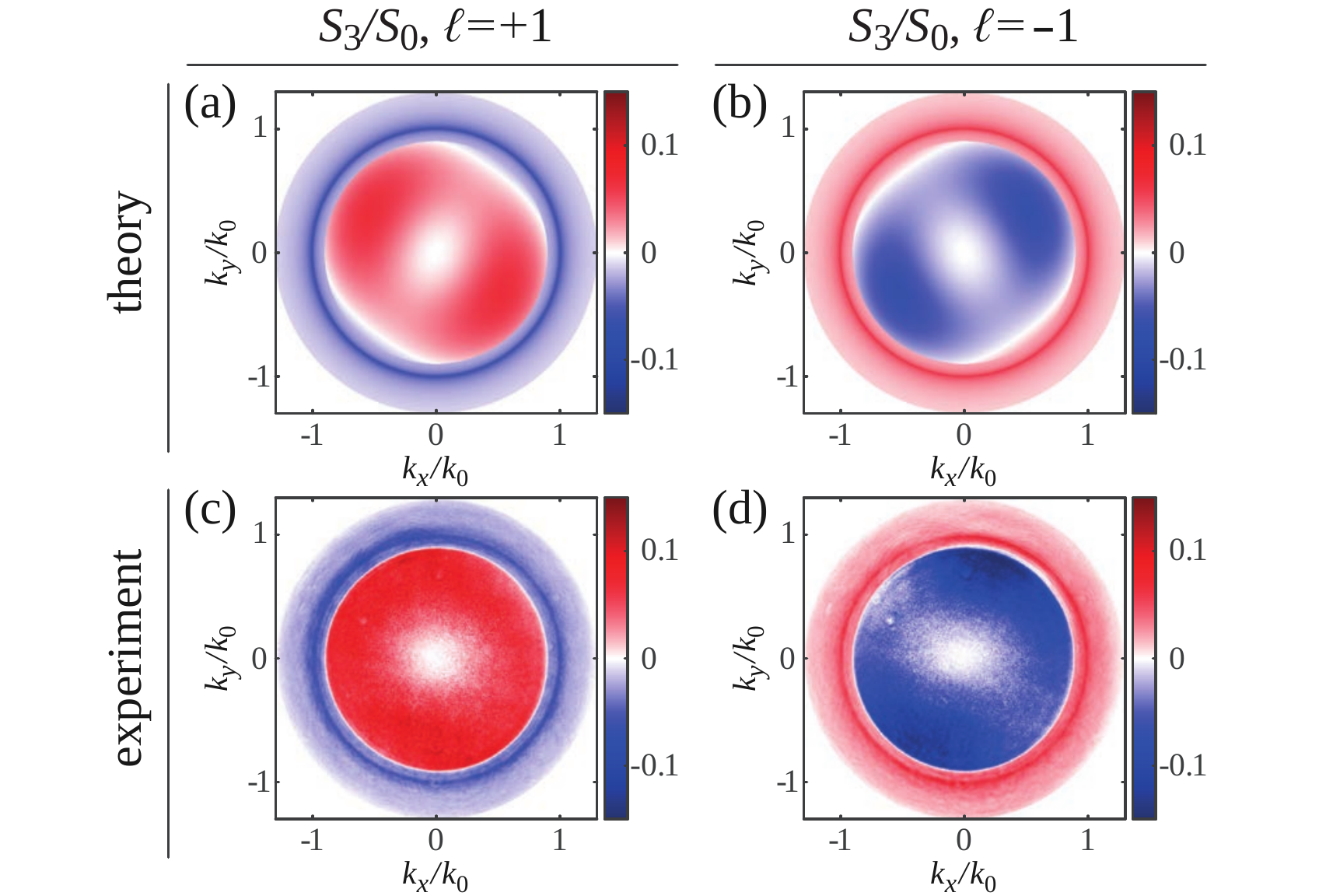}
  \caption{Theoretically calculated and experimentally measured back focal plane images of the second microscope objective. The colormap corresponds to the third Stokes parameter $S_3$, normalized by the maximum of $S_0$. (a) and (c) show the case of an azimuthal index $\ell=+1$ of the incoming LG mode, whereas (b) and (d) show results for $\ell=-1$.}
  \label{fig:results}
\end{figure}

Due to technical limitations, in practice it is not possible to collect and collimate the complete far-field of the lower half-space. Nevertheless, analyzing only the light with $k_\bot/k_0 \leq 1.3$ gives us sufficient information, because the amount of light emitted to higher transverse $k$-vectors is negligibly small. In Fig.~\ref{fig:results} (a) and (b), we show the theoretically calculated BFP images of the third Stokes parameter normalized by the maximum of $S_0$ for an incoming $\text{LG}_{+1}$ and $\text{LG}_{-1}$ beam, respectively. Below those images, in Fig.~\ref{fig:results} (c) and (d), we also present our measured results, showing a clear overlap to the theoretical counterparts.

To investigate the conservation of helicity, similar as we did it in Fig.~\ref{fig:helicity}, we look at the average helicity in certain angular regions, only restricting the highest possible transverse $k$-vector to be within the the numerical aperture (NA$_2=1.3$) of the utilized immersion-type MO. In Table \ref{tab:results} we list the theoretical and experimental results of the average helicity in those regions. Again we see a good correspondence between our theoretical predictions and the experimental findings. Most importantly we notice that at $\lambda_{d,s}=685$\,nm the total helicity is very close to zero  (see table entries for angular ranges within [0, 1.3]), proving the global conservation of helicity for a dual-symmetric scatterer. The reason for the small residual helicity origins in the discarded light emitted outside of the measured angular range. Also clearly visible from our results is the influence of the orbit-to-spin coupling upon scattering. Although the total helicity is unaffected, after the interaction of the linearly polarized LG beam with the nanoparticle, a significant amount of light is circularly polarized when looking at specific regions in the far-field.

\newcolumntype{C}[1]{>{\centering\arraybackslash}m{#1}}
\begin{table}
	\centering
	\caption{Theoretical and experimental results of the average helicity for certain regions of transverse $k$-vectors and a wavelength of 685\,nm.}\label{tab:results}	
	\begin{tabular}{>{\centering}m{0.22\columnwidth} C{0.15\columnwidth} C{0.22\columnwidth} C{0.22\columnwidth}}
	\hline 
	$k_\bot/k_0$ \hspace{3cm} region & $\ell$ & $S_3/S_0$ \hspace{3cm} theory & $S_3/S_0$ experiment \\
	\hline 
	$[0, 1.3]$   & ~1 & ~0.005  & ~0.060 \\
	$[0, 1.3]$   & -1 & -0.005  & -0.074 \\
	$[0, 0.9]$   & ~1 & ~0.045  & ~0.124 \\
	$[0, 0.9]$   & -1 & -0.045  & -0.149 \\	
	$[0.9, 1.3]$ & ~1 & -0.916  & -0.893 \\
	$[0.9, 1.3]$ & -1 & ~0.916  & ~0.868 \\
	\hline 
	\end{tabular} 
\end{table}


\section{Conclusion}
In conclusion, we have investigated orbit-to-spin angular momentum conversion upon scattering of a focused linearly polarized Laguerre-Gaussian beam by a spherical high-index dielectric nanoparticle. By tight focusing of a linearly polarized Laguerre-Gaussian beam, we create spatially varying distribution of helicity density in the focal plane. Placing a dipolar scatterer in the focal plane to locally manipulate the helicity density paves the way for manipulations on the total helicity properties of our system. These manipulations were shown to affect the spin angular momentum of the beam and the total helicity. Specifically, a dual-symmetric scatterer positioned on the optical axis resulted in the emission of purely circularly polarized light with a handedness depending on the orbital angular momentum of the incident beam, although the initial beam itself features zero helicity and zero spin angular momentum. For the case of a dual dipolar scatterer, we also demonstrated theoretically as well as experimentally the conservation of the total helicity of the interference between incident and scattered light. There, a higher index dielectric substrate allowed us to separate the transmitted far-field of the excitation beam from the purely circularly polarized scattered light, facilitating orbit-to-spin angular momentum conversion in specific angular regions. Our work provides an insight into local and global properties of helicity conservation theorems and emphasizes the role of duality symmetry in optics.\\
\begin{acknowledgments}
We gratefully acknowledge fruitful discussions with Martin Neugebauer.
\end{acknowledgments}

\bibliography{bib}
\end{document}